\documentclass[prb,twocolumn,showpacs,floatfix,superscriptaddress,unsortedaddress]{revtex4}

\usepackage{revsymb}
\usepackage{amsmath,amssymb}
\usepackage{graphicx,epsfig,subfigure,overpic,color}
\usepackage{mathrsfs}
\usepackage{multirow}
\usepackage{hyperref}
\usepackage{url}
\usepackage{color}
\usepackage{framed}
\usepackage[sort&compress]{natbib}
\usepackage{appendix}
\usepackage{pifont}
\usepackage{comment}
\usepackage{fancyhdr}
\usepackage{listings}
\usepackage{times}
\def \beq {\begin{equation}}
\def \edq {\end{equation}}
\def \bes {\begin{subequations}}
\def \eds {\end{subequations}}
\def \beqn {\begin{equation*}}
\def \edqn {\end{equation*}}

\def \dag {\dagger}

\def \up {\uparrow}
\def \down {\downarrow}

\def \sm {\sigma}

\def \veps {\varepsilon}
\def \weps {\widetilde{\varepsilon}}
\def \wD {\widetilde{D}}

\def \calh {{\cal{H}}}

\def \calf {{\cal{F}}}

\def \cals {{\cal{S}}}

\newcommand\orb{{\mathrm{orb}}}%

\begin{document}

\title{Kramers polarization in strongly correlated carbon nanotube quantum dots}
\author{Jong Soo Lim}
\affiliation{Departament de F\'{i}sica, Universitat de les Illes Balears,
  E-07122 Palma de Mallorca, Spain}
\author{Rosa L\'opez}
\affiliation{Departament de F\'{i}sica, Universitat de les Illes Balears,
  E-07122 Palma de Mallorca, Spain}
\affiliation{Institut de F\'{i}sica Interdisciplinar i de Sistemes Complexos
  IFISC (CSIC-UIB), E-07122 Palma de Mallorca, Spain}
\author{Gian Luca Giorgi}
\affiliation{Institut de F\'{i}sica Interdisciplinar i de Sistemes Complexos
  IFISC (CSIC-UIB), E-07122 Palma de Mallorca, Spain}
\author{David S\'anchez}
\affiliation{Departament de F\'{i}sica, Universitat de les Illes Balears,
  E-07122 Palma de Mallorca, Spain}
\affiliation{Institut de F\'{i}sica Interdisciplinar i de Sistemes Complexos
  IFISC (CSIC-UIB), E-07122 Palma de Mallorca, Spain}
\date{\today}

\begin{abstract}
Ferromagnetic contacts put in proximity with carbon nanotubes induce spin and orbital polarizations. These polarizations affect dramatically the Kondo correlations occurring in quantum dots formed in a carbon nanotube,
inducing effective fields in both spin and orbital sectors.  As a consequence, the carbon nanotube quantum dot spectral density shows a four-fold split SU(4) Kondo resonance. Furthermore,  the presence of spin-orbit interactions leads to the occurrence of an additional polarization among time-reversal electronic states (polarization in the time-reversal symmetry or Kramers sector). Here, we estimate the magnitude for the Kramer polarization in realistic carbon nanotube samples and find that its contribution is comparable to the spin and orbital polarizations. The Kramers polarization generates a new type of effective field that affects only the time-reversal electronic states. We report new splittings of the Kondo resonance in the dot spectral density which can be understood only if Kramers polarization is taken into account. Importantly,
we predict that the existence of Kramers polarization can be experimentally detected by performing nonlinear differential conductance measurements. We also find that, due to the high symmetry required to build SU(4) Kondo correlations, its restoration by applying an external field is not possible in contrast to the compensated SU(2) Kondo state observed in conventional quantum dots. 
\end{abstract}
 \maketitle

\section{Introduction}
Low dimensional carbon allotropes show 
extraordinary electrical and magnetic properties~\cite{generalref1,generalref2} with
potential applications in the construction of nanoelectromechanical devices,~\cite{Park} quantum
computation~\cite{quantumcomputation} and spintronics.~\cite{spintronics}
They are viewed as prime candidates to replace traditional silicon-based electronics in a new era of field-effect transistors.~\cite{fieldeffect,fieldeffect2}~For most of semiconductor
spintronic devices their functionality is limited by short spin relaxation lifetimes due to the 
hyperfine interaction.~\cite{hyperfine} Carbon-based materials overcome this obstacle and offer a nearly nuclear spin-free environment presenting very
long spin relaxation lifetimes~\cite{spinrelaxationcarbon}. As a consequence, spin injection and detection
in carbon nanotubes is carried out very efficiently.~\cite{ferrocontacts} 
Another remarkable characteristic of carbon nanotubes
 is that they can be contacted to different
types of reservoirs such as ferromagnetic~\cite{spintronics,ferrocontacts} or
superconducting materials.~\cite{supercontacts}

A metallic or semiconductor nanotube is created by
rolling up a layer of graphene along a 
given direction into a hollow cylinder.~\cite{generalref2}
Then, the carbon nanotube description is based on graphene's band structure.
Graphene consists of a monoatomic monolayer of carbon atoms 
assorted on a honeycomb lattice and its peculiarity arises from its energy dispersion relation consisting of two independent cone-shaped valleys that touch at the $K$ and $K'$ points. The low-energy excitations are chiral
fermions that obey a
massless Dirac equation with a velocity $v_F\approx 8 10^5$m/s.
Therefore, many relativistic physical phenomena, such as the
Klein paradox or the \emph{Zitterbewegung} effect can be observed
in graphene but at much lower energies.~\cite{generalref2,klein} In a carbon nanotube the wave-vector perpendicular to the tube becomes quantized, then $k_\xi=\tau k_0$ [$k_0=1/(3 R)$, $R$ being the tube radius] corresponds to the
$K$ and  $K'$ Dirac points with $\tau=\pm$,~\cite{generalref1,kan97} where $\tau$ is usually termed \emph{isospin} or \emph{valley index}. Semiclassically, the two values of
$k_\xi$ are seen as clockwise $\circlearrowright$  
and counterclockwise $\circlearrowleft$ orbits around the tubular axis that define an orbital magnetic moment ($\mu_{\rm orb}=e v_F R/2$) 10-20 times larger
than the electronic spin magnetic moment.\cite{Minot04} As a
result, the electronic states in a carbon nanotube are the four-fold degenerate eigenkets $|\tau,\sigma\rangle$, i.e., $\{|+,\uparrow\rangle$, $|+,\downarrow\rangle$, $|-,\uparrow\rangle$, $|-,\downarrow\rangle\}$. 

\begin{figure}
\centering \includegraphics[width=8cm]{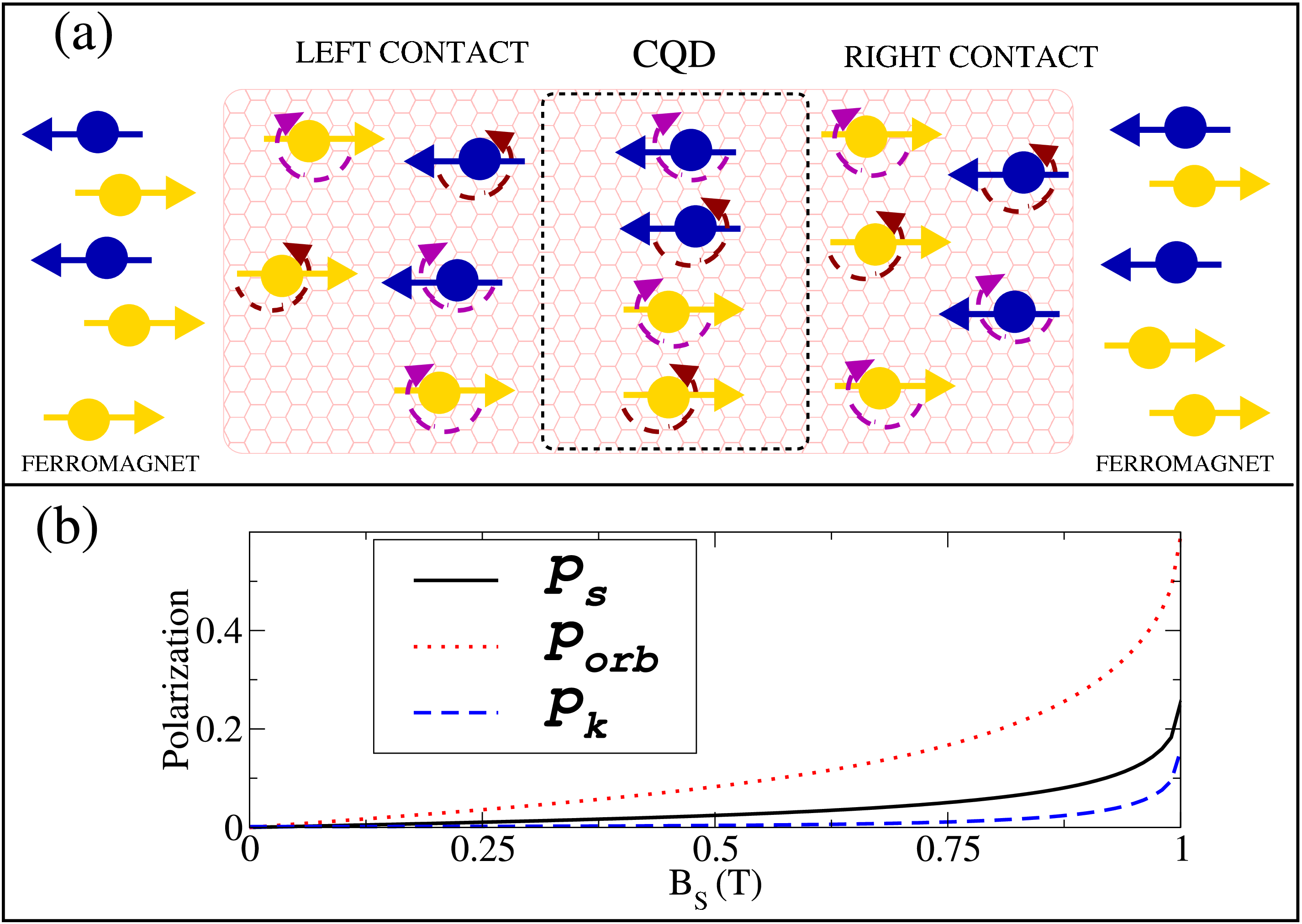} 
\caption{(Color online) (a) Schematics of the carbon nanotube quantum dot system coupled to ferromagnetic
contacts in the presence of spin-orbit interaction. Ferromagnetic electrodes, with a polarization parallel along the tube axis, are attached to the carbon nanotube. Majority (minority) spins are depicted in yellow (blue).
The ferromagnets induce a stray field ($B_{S}$) parallel to the tube axis,
which polarizes the carriers in the nanotube regions closest to the electrodes.
The polarization is not only induced in the spin (denoted with an arrow) and orbital 
(denoted with clockwise or counterclockwise orbits) sectors
but also between time-reversal electronic states (the Kramers polarization)
due to the presence of spin-orbit interactions inside the tube.
Next, a quantum dot is created in a central segment of the nanotube using a back gate.
As a result, in our model the leads coupled to the dot contain electrons
which, quite generally, are polarized carriers in the spin, orbital, and Kramers sectors.
Note that in the figure we depict the four allowed energy states in the dot.
(b) Spin, orbital and Kramers polarization  versus the stray field are shown  for $R=2.5$ nm,
$\Delta_{\rm so}=0.5 E_g$, $\mu=10 E_g$  (gap energy $E_g=\hbar v_F/3R=70$ meV),  and $\mu_{\rm orb}\approx 10\mu_B$
(values taken from the experiment reported in Ref.~\onlinecite{SOexperiment}).}
\label{scheme}
\end{figure}

Carbon nanotubes exhibit Coulomb blockade effects that imply the formation of a quantum dot
inside the tube.~\cite{SET} The dot can be created between Schottky barriers formed at the interface
between the nanotube and the metallic electrode and a transition between ballistic (Fabry-Perot)
and low transparency (Coulomb blockade) regimes is observed.\cite{transition}
Conductance measurements in carbon nanotube quantum dots reveal a four-fold
degeneracy which follows from the four-electron periodicity of the electron addition pattern. 
Increasing the transparency of the tunnel barriers between the carbon nanotube
and the contacts favors the observation of cotunneling and Kondo effects.\cite{KondoCNTs}
In highly symmetric carbon nanotube quantum dots,
electronic spin and orbital degrees of freedom are conserved when they tunnel from the contacts to the four-fold
degenerate single -particle dot states. This situation takes place when the metallic contacts
are deposited on top of the nanotube and the dot is formed between
the barriers but there still exists the probability for nanotube-nanotube tunneling events
between adjacent sections of the tube.\cite{KondoSU4CNTs_exp}
At very low temperatures ($T\ll T_K$, with $T_K$ the Kondo temperature)
and when the dot contains an odd number of electrons, the \emph{SU(4) Kondo effect}~\cite{KondoSU4}
can take place in which high-order tunneling events with
simultaneous flips in the spin and orbital sectors entangle both degrees of
freedom.~\cite{KondoSU4CNTs_exp,KondoSU4CNTs}

Clearly, the presence of magnetic interactions, such as external magnetic 
fields or spin-orbit interactions, affect dramatically 
the formation of highly-symmetric Kondo states.\cite{KondoSO,lip2}
In fact, the importance of the spin-orbit
interaction has been recently proved in ultra-clean carbon nanotubes
by Kuemmeth {\it et al.}\cite{SOexperiment} There, the
spin-orbit interaction due to the nanotube curvature~\cite{socnt} appears as a
splitting of magnitude $\Delta_{\rm so}$ between time-reversal pair states
termed Kramers degenerate states, i.e.,
$\{|+,\uparrow\rangle,|-,\downarrow\rangle\}$ and  $\{|+,\downarrow\rangle,|-,\uparrow\rangle\}$.
On the other hand, ferromagnetic electrodes are shown to destroy the Kondo resonance when the magnetic
moments are aligned in a parallel configuration.\cite{pasupathy}
Hence, it is natural to ask to what extent the transport properties of an ultra-clear carbon nanotube
are altered in the presence of both ferromagnetism and spin-orbit interaction.

A fundamental effect of ferromagnetic contacts consists of inducing stray fields in attached nanostructures.
Magnetic stray fields from patterned ferromagnetic structures are considered useful for spin manipulation due to their advantage of having rather 
high magnetic fields confined into small length scales. Different approaches have been taken to characterize and detect magnetic stray fields.\cite{stray} 
The influence of the magnetic stray fields on the spin-states in semiconductor materials has been investigated by photoluminescence,\cite{photo} 
spin-flip light scattering,\cite{stray2}
and cathodoluminiscense in semiconductor quantum wells.\cite{stray3} Our results show that due to the combined action of spin-orbit interaction and 
stray fields a new type of polarization between time-reversal electronic states is generated. Our goal in this work is to investigate the influence of all 
three polarizations (spin, orbital and Kramers) in the formation of the SU(4) Kondo state in a carbon nanotube quantum dot. 

This article is organized as follows: In Sec.~\ref{modelcnt}, we revise the theoretical model for an infinite nanotube. 
Section~\ref{polarizations} contains the calculation for the spin, orbital and Kramers polarizations in the presence of 
ferromagnet stray-fields and spin-orbit interaction. In Sec.~\ref{exchange}, we discuss our system's model Hamiltonian 
whereas in Sec.~\ref{effective} we calculate the effective fields using the scaling procedure and the projection method 
from an effective Hamiltonian. Section~\ref{results1} presents the numerical results for the calculated effective fields 
in the absence and in the presence of spin-orbit interactions. In Sec.~\ref{results2}, we show the calculation of the 
spectral density of states of a carbon nanotube quantum dot in the presence of ferromagnetism and spin-orbit interactions. 
Finally, we conclude in Sec.~\ref{conclusions} summarizing our main findings.

\section{Model of an infinite carbon nanotube}\label{modelcnt}
Carbon nanotubes are formed by wrapping a one-atom-thick layer of graphene into a cylinder. 
Carbon atoms in graphene are arranged in a two-dimensional hexagonal lattice with two carbons 
in the unit cell. This honeycomb lattice  can be considered as a combination of two overlaying triangular
 sub-lattices, $A$ and $B$. The primitive lattice vectors are $a_1=a_0(1,0)$ and $a_2=a_0(1/2, \sqrt{3}/2)$  with a lattice space $a_0=2.46$~\AA . 
 The manner in which the graphene sheet is rolled up to create a nanotube is represented by the chiral vector, $C=m_1 a_1 + m_2 a_2$, $(m_1,m_2)$ 
 being integer numbers. From the polar representation 
 of $C=2\pi R e^{i\varphi}$ one determines the nanotube radius $R=|C|/2\pi$ and its chirality $\varphi$. 
 Neglecting curvature effects and the spin-orbit interaction, the Hamiltonian for the non equivalent valleys of the 
 Brillouin zone $\textbf{K}=(2\pi/a_0)(1/3,1/\sqrt{3})$, and  $\textbf{K}^\prime=(2\pi/a_0)(-1/3,1/\sqrt{3})$ is
\begin{equation}
\mathcal{H}_{\rm CNT}= \hbar v_F (\tau k_\xi s_1 + k_\theta s_2)\,.
\end{equation}
where $v_F$ is the Fermi velocity, $\tau$ takes the value $\tau= +1$ for the  $\textbf{K}$-valley and  $\tau=-1$ 
for the  $\textbf{K}^\prime$-valley and $s_i$ ($i=1,2,3$) are the Pauli matrices corresponding to the sub-lattice space.
$k_\xi$ is the wave-vector component along the nanotube circumference and $k_\theta$ corresponds to component 
of the wave-vector along the tubular axis.
The eigenvalues of $\mathcal{H}_{\rm CNT}$ are
\begin{eqnarray}
E_{k_\xi,k_\theta}= \pm \hbar v_F \sqrt{k_\xi^2 + k_\theta^2}\,,
\end{eqnarray}
and its eigenstates read
\begin{eqnarray}
\Psi^{\textbf{K(K)}^\prime}_{k_\xi,k_\theta}(\rho)=\frac{1}{\sqrt{4\pi}} e^{i\textbf{K(K)}^\prime \cdot \rho} e^{i\left(k_\xi R \phi+ k_\theta \eta \right)}\left(\begin{array}{c} {b}_{k_\xi,k_\theta}^{\textbf{K(K)}^\prime} \\ 1 \end{array} \right)\,,
\end{eqnarray}
where 
\begin{align}
b_{k_\xi,k_\theta}^\textbf{K}&=\pm \frac{ k_\xi-i k_\theta}{\sqrt{k_\xi^2 + k_\theta^2}}
\,,\\ \nonumber
b^{\textbf{K}^\prime}_{k_\xi,k_\theta}&=\mp \frac{k_\xi+i k_\theta}{\sqrt{k_\xi^2 + k_\theta^2}}\,,
\end{align}
where $R\phi$ denotes the azimuthal direction of the nanotube and $\eta$ corresponds to the tube direction with a position vector $\rho=(R\phi\cos\varphi-\theta \sin\varphi, R\phi\sin\varphi+\theta \cos\varphi )$.
Now we roll up the sheet of graphene to create an infinite carbon nanotube, which is equivalent to impose periodic boundary conditions along the nanotube circumference: $\Psi (r+C)=\Psi(r)\rightarrow k_\xi+\textbf{K}(\textbf{K}^{'})=2\pi p$, with $p$ an  integer. The wave-vector along the nanotube circumference becomes quantized as $k_\xi= (p-\tau \nu/3)/R$. $\nu$ can take the values $0,\pm 1$ depending on the chirality of the nanotube. The chirality affects the conductance of the nanotube. Thus, a nanotube is considered metallic if the value $m_1 - m_2$ is divisible by three (and then $\nu=0$). Otherwise, the nanotube is semiconducting (for $\nu=\pm 1$). Hereafter, we treat the case of a semiconductor nanotube. 

\section{Spin, orbital and Kramers polarizations}\label{polarizations}

As mentioned before, ferromagnetic electrodes attached to carbon nanotubes induce magnetic stray fields that produce spin and orbital polarizations. For simplicity, we consider collinear ferromagnets with an easy-axis parallel to the nanotube axis. The influence of the stray field (hereafter, denoted by $B_S$) on the nanotube Hamiltonian is two-fold: (i) it generates a Zeeman term $\Delta_Z \sigma$ ($\sigma=\pm 1/2$ is the spin projection along the chiral vector) with an associated energy splitting $\Delta_Z=\hbar\omega_Z= |e| g B_S / 2 m_0 c $ ($e$ electron charge, Land\'e factor $g=2$ for carbon atoms, $m_0$ effective electron mass, and $c$ speed of light); and (ii) it produces an Aharanov-Bohm flux $\Phi_{\rm AB}=\pi R^2  B_S$ that threads the carbon nanotube and modifies the boundary condition ($p\rightarrow p+\Phi_{\rm AB}/\Phi_0$) for the wave-vector along the tube circumference: 
\begin{equation}
k_{\xi}\rightarrow p-\frac{\tau\nu}{3R} + \frac{\Phi_{\rm AB}}{\Phi_0 R}\,,
\end{equation}
with $\Phi_0=h/e$ as the flux quantum.  For the lowest sub-band, we have $p=0$ and the energy gap in a semiconductor nanotube ($\nu=\pm 1$) is  $E_{g}=\hbar v_F/3R$.

With all these ingredients, the energy dispersion relation for the lowest sub-band ($p=0$) of an infinite semiconductor carbon nanotube (hereafter, we take $\nu=1$) in the presence of a magnetic stray-field $B_S$ reads (to shorten the notation we set $k_\theta\rightarrow k$),
\begin{eqnarray}\label{dispersion1}
E(k)= \pm \hbar v_F \sqrt{\left(\frac{\tau}{3R}+ \frac{\Phi_{\rm AB}}{\Phi_0 R}\right)^2+ k^2}+ \hbar\omega_Z \sigma\,.
\end{eqnarray}

The density of states per unit length is calculated from  Eq.~(\ref{dispersion1})
\begin{equation}\label{dosinfinitecnt}
\Theta_{\tau\sigma}(E)=\begin{cases} \frac{1}{\hbar v_F}
\frac{|E_\sigma|}{\sqrt{(E_\sigma)^2-(\tau E_{g}+ \mu_{\rm orb}B_S)^2}}&
\text{if $E_\sigma >|\epsilon_\tau$}|\,,\\
0&
\text{if $E_\sigma <|\epsilon_\tau $}|\,,
\end{cases}
\end{equation}
where  $E_{\sigma}=E-\sigma\hbar\omega_Z$, and  $\epsilon_\tau=\tau E_{g}+ \mu_{\rm orb} B_S$ with $\mu_{\rm orb} B_S=\hbar v_F\Phi_{\rm AB}/\Phi_0 R$ . Notice that the density of states depends on the orbital $\tau$ and spin $\sigma$ quantum numbers.

From the density of states, $\Theta_{\tau\sigma}(E)$, the  spin ($p_s$) and orbital polarizations ($p_{\rm orb}$) can be obtained.~\cite{lip2} First, we calculate the spin and orbital populations  by integrating on energy the density of states [Eq.~(\ref{dosinfinitecnt})],
\begin{equation}
n_{\tau\sigma}=\int _{\epsilon_{\tau}+\hbar\omega_Z}^{\mu}\Theta_{\tau\sigma}(E) dE\,.
\end{equation} 
with $\mu$ as the electrochemical potential which can be tuned with a nearby gate voltage.\cite{Devoret97} Then, the total population of carriers per spin $\sigma$ becomes
\begin{equation}\label{pospin}
n_\sigma=\sum_\tau n_{\tau\sigma}\,,
\end{equation} 
and the total population of carriers per orbital $\tau$ is given by
\begin{equation}\label{poor}
n_\tau=\sum_\sigma n_{\tau\sigma}\,.
\end{equation} 
These populations [Eq.~(\ref{pospin}), and Eq.~(\ref{poor})] define the spin polarization,
\begin{eqnarray}
p_s=\sum_\tau  \frac{n_{\tau\downarrow}-n_{\tau\uparrow}}{n_{+\uparrow}+ n_{+\downarrow}+n_{-\uparrow}+n_{-\downarrow}}\,,
\end{eqnarray}
and the orbital polarization,
\begin{eqnarray}
p_{\rm orb}=\sum_\sigma  \frac{n_{+\sigma}-n_{-\sigma}}{n_{+\uparrow}+ n_{+\downarrow}+n_{-\uparrow}+n_{-\downarrow}}\,.
\end{eqnarray}

We now include the effect of spin-orbit interactions due to the nanotube curvature.~\cite{SOexperiment,socnt,KondoSO}
The spin-orbit interaction enters as a spin-dependent topological flux $\sigma\Phi_{\rm so}$
that additionally shifts $k_\xi$,\cite{socnt}
\begin{equation}
k_\xi\rightarrow k_\xi -\sigma\frac{\Phi_{\rm so}}{\Phi_0 R}\,,\,\,\,\,\,\, {\rm with}\,\,\,\,  \frac{\Phi_{\rm so}}{\Phi_0}=\frac{\Delta_{at}}{12\epsilon_{\pi\sigma}}\left( 5+ 3 \frac{V_{pp}^\sigma}{V_{pp}^\pi}\right)\,,
\end{equation}
where $\Delta_{at}$ is the energy splitting due to the atomic spin-orbit coupling in the $p$ bands, $\epsilon_{\pi\sigma}$ is the energy splitting of the $p$ and $s$ bands in graphene,  $V_{pp}^{\pi}$, and $V_{pp}^{\sigma}$  are the hopping elements within these bands. The energy splitting associated with the spin-orbit interaction is  $\hbar v_F \Phi_{\rm so}/(\Phi_0 R)=\Delta_{\rm so}$ and the energy dispersion relation including spin-orbit interaction becomes\cite{noteSO},
\begin{equation}\label{dispersion2}
E(k)=\pm \hbar v_F \sqrt{\left(\frac{\tau}{3R}+ \frac{\Phi_{\rm AB}}{\Phi_0 R}-\frac{\Phi_{\rm so}\sigma}{\Phi_0 R}\right)^2+ k^2}+ \hbar\omega_Z \sigma\,.
\end{equation}
Notably, the spin-orbit interaction  modifies the nanotube density of states and induces a polarization in the presence of a magnetic field between electronic time-reversal pair states  defined as 
\begin{equation}
p_k=\frac{n_{+\uparrow}+n_{-\downarrow}-n_{+\downarrow}-n_{-\uparrow}}{n_{+\uparrow}+n_{-\downarrow}+n_{+\downarrow}+n_{-\uparrow}}\,,
\end{equation}
with a modified density of states by the presence of $\Delta_{\rm so}$
\begin{equation}\label{dosinfinitecntso}
\Theta_{\tau\sigma}(E)=\begin{cases}
\frac{1}{\hbar v_F}\frac{|E_\sigma|}{\sqrt{(E_\sigma)^2-(\tau E_{g}+ \mu_{\rm orb}B_S + \Delta_{\rm so}\sigma)^2}}&
\text{if $E_\sigma >|\epsilon_{\tau\sigma}|$}\,,\\
0&
\text{if $E_\sigma <|\epsilon_{\tau\sigma}|$}\,,
\end{cases}
\end{equation}
where
\begin{equation}
\epsilon_{\tau\sigma}=\tau E_{g}+\mu_{\rm orb}B_S +\Delta_{\rm so}\sigma\,,
\end{equation}
Notice that if $\Delta_{so}=0$, then $p_k$ vanishes. However, even if the density of states shows the coupling between the spin and orbit quantum numbers, it might occur that the generated polarization in the Kramers sector is very small. Our results demonstrate that this, in fact, is not the case for realistic carbon nanotubes. We calculate the three types of polarization present in a nanotube attached to ferromagnetic electrodes. For the numerical simulation we have utilized the parameters reported in Ref.~[\onlinecite{SOexperiment}]: $R=2.5$ nm, $\Delta_{\rm so}=$0.37 meV, $\mu_{\rm orb}\approx$ 12 $\mu_s$.  Figure~1(b) shows the three polarizations. It is worth noting that detectable orbital and Kramers polarizations, of the order of 10-20 per cent, can be achieved at moderate stray fields around $B_S=1$~T. 

\section{Hamiltonian} \label{exchange}

In the previous section, we discussed the existence of a new type of polarization among the time-reversal electronic states of a very long carbon nanotube attached to ferromagnets in the presence of spin-orbit interaction. Moreover, we proved that this polarization is not negligible and it can reach finite values at moderate stray fields. In this section, we address a different problem, namely, the detection of such a polarization in the transport properties of a carbon nanotube quantum dot. 
A quantum dot can be created in a nanotube by applying a backgate potential onto a short segment of the nanotube in order to produce a depleted region of electrons. Additionally, other mechanisms such as lattice-mismatch or the presence of defects can create quantum dots but in a very uncontrollable manner. 

As we discussed in Sec.~\ref{polarizations}, due to the presence of spin-orbit interaction, in addition to spin and orbital polarizations a new polarization between time-reversal electronic states arises.  Therefore, in our model for the leads we fully take into account the three types of polarization. For the quantum dot, we use a four-fold degenerate Anderson model that includes spin-orbit effects. See Fig. 1(a) for a sketch of the system. Then, the Hamiltonian reads,
\begin{multline}\label{hamiltonian}
\mathcal{H}=\sum_{\alpha=L/R,k,\tau\sm} E_{k\tau\sm}^\alpha c_{\alpha k \tau\sm}^{\dagger}c_{\alpha k\tau\sm} +\sum_{\tau,\sm} \veps_{d\tau\sm} d_{\tau\sm}^{\dag} d_{\tau\sm} \\
+ U \sum_{\tau\sm \ne \tau'\sm'} n_{d \tau\sm}n_{d \tau'\sm'}+\sum_{\alpha=L/R,k,\tau\sm} \left( V_{\alpha} c_{\alpha k \tau\sm}^{\dag} d_{\tau\sm} + h.c.\right).
\end{multline}
The first term describes the leads where $c_{\alpha k\tau\sm}$ destroys an electron in the reservoir $\alpha=L/R$ with quantum numbers $k\equiv k_\theta,\tau,\sm$. The dot operator $d_{\tau\sigma}$  annihilates an electron of orbital mode $\tau$ and spin $\sigma$ on the dot. $U$ denotes the charging energy, $n_{d \tau\sm}=d_{\tau\sigma}^\dagger d_{\tau\sigma}$ the dot occupation and $V_\alpha$ the tunneling amplitude for the $\alpha$ barrier. For the quantum dot region, the condition for the quantization of $k$ [see Eq.~(\ref{dispersion2})] leads to the following expression for the
single-particle levels:
\begin{equation}
\veps_{d\tau\sm} = \veps_{d} + \sm\tau \frac{\Delta_{\rm so}}{2} + \tau \mu_{\rm orb} B + \sm g \mu_{s} \frac{B}{2}\,,
\end{equation} 
where $B$ denotes any external magnetic field applied  to the quantum dot.

In the absence of magnetic interactions and at very low temperatures, this Hamiltonian exhibits the celebrated SU(4) Kondo effect where simultaneous fluctuations in the orbital and spin sectors  build a highly symmetric correlated state.~\cite{KondoSU4CNTs} 

\section{Effective fields}\label{effective}

When the orbital degree of freedom is absent, as it occurs for example
in semiconductor quantum dots created by gating two-dimensional electron gases,
the observed Kondo effect has SU(2) symmetry. The problem of ferromagnetic contacts
attached to such quantum dots has been extensively studied in  Refs.~\onlinecite{ferronrg1,ferronrg2}.
There, it was demonstrated that charge fluctuations lead to a different renormalization of the dot energy levels depending on the spin direction.  In this fashion, the effect of ferromagnetic contacts is seen as an \emph{effective field} $B_{\rm eff}$ that breaks the spin degeneracy on the dot. Recently, similar effective fields have been analyzed in spin-orbit quantum dots inserted in Aharanov-Bohm interferometers.\cite{ring1,ring2}
In these cases, the Kondo resonance can be restored by applying an appropriate external magnetic
field $B$ which fulfills the condition $B_{\rm eff}+B=0$.
In our nanotube system attached to ferromagnetic contacts,
spin, orbital and Kramers polarizations are present.
As a consequence, we expect {\em three} effective fields in the spin, orbital, and Kramers sectors.

To gain some physical intuition we consider a simplified model for the leads.
We consider their density of states to be energy independent, which we parametrize
using $p_s$, $p_{\rm orb}$, and $p_{k}$:
\begin{eqnarray}\label{densityleads}
\Theta^{L/R}_{\sigma\tau}(E) = \Theta^{L/R} \left(1+\sigma p_s +\tau p_{\orb} +\tau\sm p_k\right ),
\end{eqnarray}
where $\Theta^{L/R}=1/2D_0$ with $2D_0$ the lead bandwidth. In what follows,
we use Eq.~(\ref{densityleads}) for the leads' density of states
and take the values of  $p_s$, $p_{\rm orb}$, and $p_{k}$ from Fig.~\ref{scheme}.

In order to see the effect of $p_s$, $p_{\rm orb}$, and $p_{k}$ on the dot spectrum
we apply the scaling technique to the dot level when the energy cutoff $\wD$ is reduced from $D_0$:\cite{hal} 
\begin{equation}\label{scaling}
\frac{d\veps_{\tau\sm}}{d\ln \wD}=\frac{-1}
{\pi}\sum_{\substack{\alpha\\(\tau'\sm')\neq(\tau\sm)}}\Gamma_{\alpha\tau'\sm'}\,,
\end{equation} 
where $\Gamma_{\alpha\tau\sm}=\pi|V_{\alpha}|^2 \Theta_{\tau\sigma}^{\alpha}$ is the tunneling rate for electrons in lead $\alpha$ with spin $\sm$ and orbital mode $\tau$. Solving the scaling equation [Eq.~(\ref{scaling})], we find that the dot level is renormalized as (hereafter, we consider symmetric tunneling amplitudes $V_L=V_R=V_0$ with $\Gamma_0=\pi V_0^2\nu_0$),
\begin{eqnarray}
&&\weps_{d\tau\sigma}=\veps_{d\tau\sigma}+\frac{\Gamma_0}{2\pi}\\ \nonumber
&&\times \sum_{\sm,\tau}\left[4-(1+\sm p_s +\tau p_{\rm orb}+\tau\sm p_k)\right]\ln \frac{D_0}{\tilde{D}}\,.
\end{eqnarray} 
This result clearly indicates a different renormalization of the quantum dot energy level depending on $\tau$ and $\sm$, 
which leads to the generation of effective fields in the spin, orbital and Kramers sectors.  

To calculate explicitly the effective fields originated by the three polarizations,
we use the projection method,\cite{hew93} where the charge fluctuations are integrated out,
and find an effective Hamiltonian,
\begin{equation}\label{proyectionmethod}
 \calh_{\rm eff} = \calh_{11} +\calh_{10}\frac{1}{E_1 - E_0}\calh_{01}+\calh_{12}\frac{1}{E_1 - E_2}\calh_{21}\,,
\end{equation} 
where $E_0$, $E_1$ and $E_2$ correspond  to the energies for empty, singly and doubly occupied quantum dot state
and the projectors are
\begin{eqnarray}
\calh_{01} &=& \sum_{\alpha, k,\tau\sm} V_{0}c_{\alpha k\tau\sm}^{\dagger} {\prod_{\tau'\sm'}}'(1-n_{d\tau'\sm'}) d_{\tau\sigma}\,,
\\\nonumber
\calh_{21}&=& \sum_{\alpha,k,\tau\sm} {\sum_{\tau'\sm'}}' V_{0}^{\ast}
d_{\tau\sm}^{\dag} n_{d\tau'\sm'} {\prod_{\tau''\sm''}}''(1 - n_{d\tau''\sm''}) c_{\alpha k\tau\sm}\,,
\end{eqnarray}
where the prime in the product appearing in $\calh_{01}$ means $\tau'\sm'\ne\tau\sm$ and the double prime in $\calh_{21}$ signifies $\tau''\sm'' \ne (\tau\sm,\tau'\sm')$. Defining the four-component spinor 
\begin{equation}
\Psi_d^\dagger = (d_{+\up}^\dagger, d_{+\down}^\dagger, d_{-\up}^\dagger, d_{-\down}^\dagger)\,,
\end{equation} 
the effective  Hamiltonian reads
\begin{equation}\label{calhrmeff}
\calh_{\rm eff} = -\sum_{\alpha, q,\sigma,\tau} \Psi_d^{\dag} \frac{1 + \tau\hat\tau_\zeta}{2} \otimes \frac{1 + \sigma\hat\sm_\zeta}{2} \Psi_d \calf_{\alpha \tau\sigma}\,,
\end{equation}
where 
\begin{eqnarray}
\calf_{\alpha q \tau\sigma} = \frac{|V_{0}|^2\left[1-f_{\alpha}(E_{ q \tau\sigma}^{\alpha})\right]}{E_{q\tau\sigma}^\alpha-\veps_{d \tau\sigma}}+ \frac{|V_{0}|^2 f_{\alpha}(E_{q\tau\sigma}^\alpha)}{E_{q\tau\sigma}^\alpha-\veps_{d\tau\sigma}-U}.
\end{eqnarray}
In Eq.~\eqref{calhrmeff}, $\hat\sm_\zeta$ and $\hat\tau_\zeta$ are the diagonal Pauli matrices for the spin and orbital sectors, respectively. It is convenient to define 
\begin{equation}
\cals_{ij} =  \frac{1}{4} \Psi_d^{\dag} (\tau_i \otimes \sm_j) \Psi_d\,,
\end{equation}
in order to give a more transparent expression for the effective Hamiltonian
\begin{multline}\label{eq:heff}
\calh_{\rm eff}= \frac{\Gamma_0}{2\pi} \sum_{\alpha\tau\sigma} \left[1+\sigma p_{s}+\tau p_{\rm orb} +\tau\sigma p_{k}\right ]\times \mathcal{B}_\alpha(\veps_{d\tau\sigma})\\
\times\left(\sigma S_{0\zeta} +\tau S_{\zeta 0} +\tau\sigma S_{\zeta\zeta}\right)= \mathbb{B}_{s} S_{0\zeta} + \mathbb{B}_{\rm orb}  S_{\zeta 0} + \mathbb{B}_{k}  S_{\zeta \zeta} \,,
\end{multline}
where (${\rm Re}[...]$ denotes the real part)
\begin{eqnarray}
&&\mathcal{B}_\alpha(x)=
\\ \nonumber
&&{\rm Re}\Biggr\{\Psi\left[\frac{1}{2}-\frac{i\left(x-\mu_\alpha\right)}{2\pi k_B T}\right] 
-\Psi\left[\frac{1}{2}-\frac{i\left(x+U-\mu_\alpha\right)}{2\pi k_B T}\right]\Biggr\}\,,
\end{eqnarray}
where $\mu_\alpha$ denotes the chemical potential for contact $\alpha$
and $\Psi$ the digamma function.
Importantly and in contrast to the spin SU(2) Kondo case
where one effective field exists only,\cite{ferronrg1,ferronrg2}
Eq.~(\ref{eq:heff}) defines {\em three} effective fields, namely, $\mathbb{B}_{s}$, $\mathbb{B}_{\rm orb}$ and $\mathbb{B}_{k}$ that act on the spin ($S_{0\zeta}$), orbital ($S_{\zeta 0}$) and Kramers sectors ($S_{\zeta\zeta}$), respectively. Clearly, the effect of these fields is to remove the spin, orbital and Kramers degeneracies. We also note that the effective fields develop in our system only to the extent
that interactions are present. For $U=0$ the effective fields vanish altogether.

\section{Results}
\subsection{Effective fields}\label{results1}
Figure 2 illustrates the dependence of the effective fields  as a function of the dot level position $\veps_d/U$ in the absence ($\Delta_{\rm so}=0$) [Fig. 2(a)] and in the presence ($\Delta_{\rm so}=4k_B T_K$) [Fig. 2(b)] of spin-orbit interactions. We observe in Fig. 2(b) that in the presence of spin-orbit interactions, the shape of $\mathbb{B}_{s}$, and $\mathbb{B}_{\rm orb}$ [cf. Fig. 2(a)] remains essentially unaltered. However, the spin-orbit interaction induces a strong Kramers field component $\mathbb{B}_{k}$. Therefore, when ultra-clean highly symmetric carbon nanotube quantum dots exhibit SU(4) Kondo physics, spin, orbital and Kramers polarizations ($p_{s}$, $p_{\rm orb}$, and $p_k$) induce effective fields in the spin, orbital and Kramers sectors ($\mathbb{B}_{s}$, $\mathbb{B}_{\rm orb}$, and $\mathbb{B}_{k}$) and consequently the SU(4) Kondo effect is destroyed. These fields induce {\em six} different splittings that correspond to the nonequivalent transitions for which $\tau,\sigma$ change to different $\tau'\sigma'$:
\begin{figure}[!h]
\centering
 \includegraphics[width=7.3 cm]{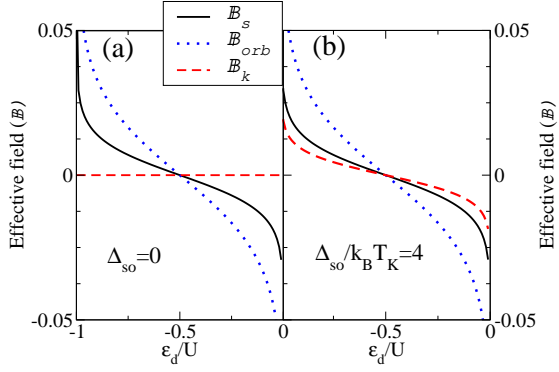} 
\caption{(Color online) Effective fields (units of $\Gamma_0$): $\mathbb{B}_{\rm s}$, $\mathbb{B}_{\rm orb}$, $\mathbb{B}_{k}$. (a) Vanishing spin-orbit interaction with $p_{s}=0.25$, $p_{\rm orb}=0.59$. (b) Nonzero spin-orbit interaction ($\Delta_{\rm so}/k_BT_K=4$) with polarizations $p_{s}=0.25$, $p_{\rm orb}=0.59$, and $p_k=0.16$. Rest of parameters: $T=T_K[\veps_d=-0.1\Gamma_0]\approx 3\times 10^{-5}D_0$ [with a Kondo temperature \cite{KondoSO} given by $T_K=D e^{-\pi|\veps_d|/3\Gamma_0}$], $\Gamma_0=0.01D_0$ and $D_0=1$.}
\end{figure}
\begin{itemize}
\item
two spin-flip \emph{intra}orbital transitions corresponding to $|\tau\sm\rangle\Leftrightarrow|\tau\bar\sigma\rangle$, induced by $\mathbb{B}_{s}$,
\item
two spin-conserved \emph{inter}orbital transitions corresponding to $|\tau,\sm\rangle\Leftrightarrow|\bar\tau,\sigma\rangle$, generated by  $\mathbb{B}_{\rm orb}$, and
\item
 two spin-flip \emph{inter}orbital transitions corresponding to $|\tau,\sm\rangle\Leftrightarrow|\bar\tau,\bar\sigma\rangle$ due to $\mathbb{B}_{k}$.
\end{itemize}
Remarkably, the higher symmetric SU(4) Kondo state lacks the compensation effect observed
in the SU(2) Kondo effect.\cite{ferronrg1,ring1,ring2}  Below, we discuss this fact analyzing the spectral function.

\subsection{Spectral density of states}\label{results2}
The dot local density of states (DOS) per spin $\sm$ and orbital mode $\tau$ is calculated from the retarded Green function as (${\rm Im}[\ldots]$ denotes the imaginary part),
\begin{equation}
A_{\tau \sm}(\omega)=- \Gamma_0 {\rm Im}\left[\mathcal{G}^r_{d\tau\sigma}(\omega)\right]\,.
\end{equation} 
Our description for the approximated dot Green function employs the slave-boson theory with non-canonical commutation relations~\cite{refmethod} generalized to account for the orbital quantum number present in carbon nanotube quantum dots. This scheme has some advantages over the standard equation of motion.~\cite{lacroix} In particular, it reproduces in dots similar values for the spin polarization than more sophisticated methods such as numerical renormalization group.~\cite{ferronrg1,ferronrg2}
Additionally, it leads to zero spin polarization at the compensation field
whereas the standard equation of motion produces an incorrect non-zero polarization.~\cite{refmethod}

Let us consider first the case of normal contacts.  When the spin-orbit interaction is absent, the low-energy DOS shows the SU(4) Kondo resonance pinned at $\omega\approx T_K$. In the presence of spin-orbit interaction, the four-fold energy dot state splits in two pairs of Kramer degenerate states with energy difference $\Delta_{\rm so}$.~\cite{KondoSO} The density of states consists of three peaks  where the central peak corresponds to the \emph{SU(2) Kramers Kondo} resonance built from high-order correlated tunneling events that involve spin-flip \emph{inter}orbital processes. The other two peaks  are related to the spin Kondo effect (spin-conserved tunneling)  and the orbital Kondo effect (orbital-conserving tunneling) in an effective magnetic field $\Delta_{\rm so}$ and therefore with identical transition energies. 

We next consider the effect of ferromagnetic contacts in the nanotube when spin-orbit effects are negligible, i.e., when $\Delta_{\rm so}=0$, as shown in Fig.~3 where the dot spectral function exhibits a four-fold split Kondo state. This is a consequence of broken spin and orbital degeneracy by the action of the effective fields $\mathbb{B}_s$ and $\mathbb{B}_{\rm orb}$. 
Here, $\mathbb{B}_s$ induces spin-flip \emph{intra}orbital transitions:
\begin{equation}
|\tau,\sm\rangle\Leftrightarrow|\tau,\bar\sigma\rangle\,,
\end{equation}
with a unique associated transition energy for both orbital modes: 
\begin{equation}
\delta_s=|\veps_{+\uparrow}-\veps_{+\downarrow}|=|\veps_{-\uparrow}-\veps_{-\downarrow}|\,.
\end{equation}
The DOS peaks originated by these type of transitions are labeled as ``1'' in Fig. 3(a) (note that each transition $\delta_i$ develops a pair of peaks in the spectral density at $\omega/\Gamma_0=\pm\delta_i$ with an associated splitting $\Delta_i=2\delta_i$).
Similarly, the two spin-conserved \emph{inter}orbital transitions:
\begin{equation} 
|\tau,\sm\rangle\Leftrightarrow|\bar\tau,\sigma\rangle\,,
\end{equation}ç
have the same transition energy for each spin orientation: 
\begin{equation}
\delta_{\rm orb}=|\veps_{+\uparrow}-\veps_{-\uparrow}|=|\veps_{+\downarrow}-\veps_{-\downarrow}|\,.
\end{equation} 
Figure 3(a) shows the peaks corresponding to spin-conserved \emph{inter}orbital transitions labeled as ``2''. Finally, the two spin-flip \emph{inter}orbital transitions 
\begin{equation}
|\tau,\sm\rangle\Leftrightarrow|\bar\tau,\bar\sm\rangle\,,
\end{equation} 
with associated peaks in the DOS are labeled as ``3'', see Fig. 3(a), are those where both spin and orbital are simultaneously changed. Here, the transition energies are:
\begin{equation}
\delta^{+}=|\veps_{+\uparrow}-\veps_{-\downarrow}|\,,\,\,\ {\rm and}\,\,\,\, \delta^{-}=|\veps_{+\downarrow}-\veps_{-\uparrow}|\,.
\end{equation}
Moreover, the magnitude of the splittings obtained from the density of states [see Fig.~3(a)] are in good agreement with those obtained using Eq.~(\ref{eq:heff}). From Fig.~3(a) we observe that $\Delta_{\rm orb}=2\delta_{\orb}\approx 0.02\Gamma_0$ and $\Delta_{s}=2\delta_{s}\approx 0.008\Gamma_0$.  These values are to be compared with the splittings $\Delta_{s}=\mathbb{B}_{s}=$ and $\Delta_{\rm orb}=\mathbb{B}_{\rm orb}$ showed in Fig.~2(a) which are of the same order ($0.005$--$0.01\Gamma_0$). 
\begin{figure}[!b]
\centering \includegraphics[width=7cm]{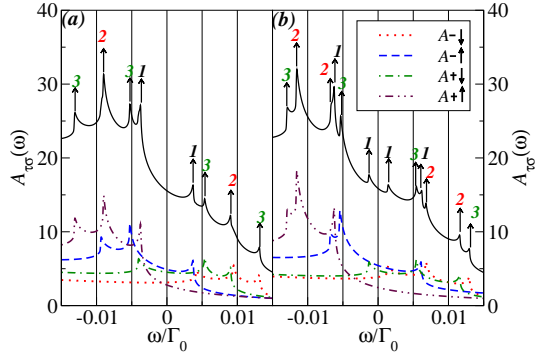} 
\caption{(Color online) Density of states $A_{\tau\sigma}(\omega)$ for a ultra-clean highly symmetric carbon nanotube quantum dot for $U=\infty$. Black solid line shows the total density of states $A(\omega)=\sum_{\tau\sigma}A_{\tau\sm}(\omega)$. (a) $\Delta_{\rm so}=0$  with $p_{s}=0.25$, $p_{\rm orb}=0.59$.  Spin-flip \emph{intra}orbital transitions (transition energy $\delta_s$) are associated to the peaks labeled as ``1'', spin-conserved \emph{inter}orbital transitions (transition energy $\delta_{\rm orb}$) correspond to peaks labeled as ``2'' and spin-flip \emph{inter}orbital transitions (transition energies $\delta^+$ and $\delta^-$) are indicated with peaks labeled as ``3'. Here,  $\delta_{s}\approx 0.004$, $\delta_{\rm orb}=0.009$, $\delta^+=0.013$, and $\delta^-=0.005$. Each transition energy $\delta_i$ generates two peaks at $\omega/\Gamma_0=\pm\delta_i$.  (b) $\Delta_{\rm so}=4T_K$ with $p_{s}=0.25$, $p_{\rm orb}=0.59$, $p_k=0.16$. Spin-flip \emph{intra}orbital transition energies $|\veps_{d+\uparrow}-\veps_{d+\downarrow}|=-0.006$, and $|\veps_{d+\uparrow}-\veps_{d+\downarrow}|=-0.0013$, spin-conserved \emph{inter}orbital transition energies $|\veps_{d+\uparrow}-\veps_{d-\uparrow}|=-0.0117\Gamma_0$, and $|\veps_{d+\downarrow}-\veps_{d-\downarrow}|=-0.007$, and spin-flip \emph{inter}orbital transition energies $\delta^+=-0.013$, and $\delta^-=-0.005$. Rest of parameters:  $\veps_d=-0.1\Gamma_0$, $\Gamma_0=0.01D_0$, $D_0=1$, $T=T_K$.}
\end{figure}
Finally,  the restoration of the SU(2) spin and the orbital Kondo effects is possible by applying an external magnetic field that cancels the splitting generated by $\mathbb{B}_{s}$ and $\mathbb{B}_{\rm orb}$. Figure 4 displays the values for which $\delta_{\rm s}$ (solid line),  $\delta_{\rm orb}$ (dotted line),  and $\delta^{\pm}$ (dashed and dot-dashed line) vanish when $B$ and $\veps_d/U$ are tuned. Notice that the spin(orbital) Kondo effect is restored whenever $\delta_{\rm s}=0$ ($\delta_{\rm orb}=0$). 

As we discussed above, when spin-orbit interactions are present a Kramers effective field $\mathbb{B}_k$ is generated. $\mathbb{B}_k$ leads to different spin-flip \emph{intra}orbital transition energies for each orbital mode 
\begin{equation}
|\veps_{d+\uparrow}-\veps_{d+\downarrow}|\neq |\veps_{d-\uparrow}-\veps_{d-\downarrow}|\,,
\end{equation} 
and different spin-conserved \emph{inter}orbital transition energies depending on the spin orientation
\begin{equation} 
|\veps_{d+\uparrow}-\veps_{d-\uparrow}|\neq |\veps_{d-\downarrow}-\veps_{d+\downarrow}|\,.
\end{equation}
Since each new transition energy develops a pair of peaks in the dot spectral function, the DOS shows {\em twelve} peaks (corresponding to the six different transition energies: two spin-flip \emph{intra}orbital, two spin-conserved \emph{inter}orbital and two spin-flip \emph{inter}orbital transitions). The new peaks in  Fig.~3(b) arising from the presence of Kramers polarization and therefore induced by the effective field $\mathbb{B}_k$ are ``\emph{the smoking gun}'' of the presence of spin-orbit interactions in carbon nanotubes exhibiting SU(4) Kondo correlations. 
Its detection would be possible by measuring the nonlinear differential conductance of an ultra-clean highly symmetric carbon nanotube quantum dot attached to ferromagnetic contacts. To some extent, for relatively weak dc source-drain bias voltages (denoted by $V_{\rm sd}$) the density of states of the nanotube can be considered as  $V_{\rm sd}$ independent. In this case, the differential conductance for the nanotube quantum dot can be approximated in the symmetric capacitive case by:
\begin{equation}
\frac{dI}{dV_{\rm sd}}\approx [A(V_{\rm sd}) + A(-V_{\rm sd}]\,.
\end{equation}
 The required energy resolution for the observation of a twelve asymmetrically located peaks (with respect to $E_F=0$) in the density of states and therefore a multi-peak structure in the differential conductance  has been already achieved in experiments with carbon nanotube quantum dots~\cite{KondoSU4CNTs} showing SU(4) Kondo physics in the presence of magnetic fields. Finally, we would like to remark that all calculations presented in this work were done using parameters extracted from experimentally available data in carbon nanotube quantum dots.~\cite{KondoSU4CNTs,SOexperiment}

\begin{figure}[!t]
\centering \includegraphics[width=7cm]{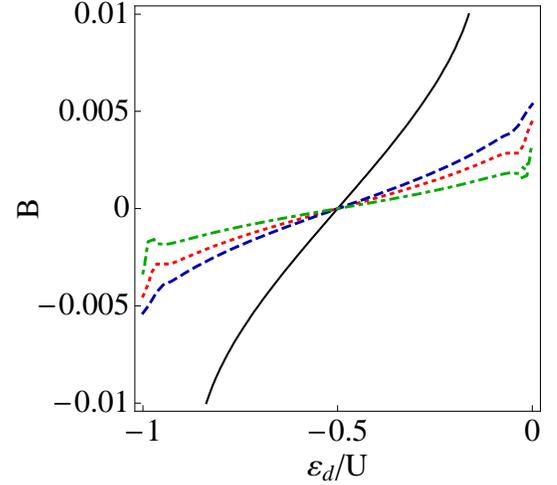} 
\caption{(Color online) Zero splitting values for the spin-flip \emph{intra}orbital transitions ($\delta_{\rm s}=0$ black solid line), spin-conserved \emph{inter}orbital transitions ($\delta_{\rm orb}=0$ red dotted line) and spin-conserved \emph{inter}orbital transitions ($\delta^{+}=0$ blue dashed line, and  $\delta^{-}=0$ green dot-dashed line) versus external field $B$ (units of $\Gamma_0$) and dot level position $\veps_d/U$. Rest of parameters: $p_{s}=0.25$, $p_{\rm orb}=0.59$, $p_k=0$, $T=T_K$, $\Gamma_0=0.01$, $D_0=1$.}
\end{figure}

\section{Conclusions}\label{conclusions}
In closing, we have shown that in carbon nanotube quantum dots attached to ferromagnetic electrodes both a spin polarization and an orbital polarization emerge. The main effect of these polarizations in a highly correlated SU(4) Kondo state is to remove the spin and orbital degeneracies and to split the Kondo resonance in eight peaks via the generation of effective fields acting on the dot. In ultra-clean carbon nanotubes quantum dots, spin-orbit coupling due to the nanotube curvature is a relevant interaction. In this case, ferromagnetic contacts induce a polarization between time-reversal electronic states. This is reflected as an emerging effective field in the Kramers sector. As a result, the Kondo peak splits into twelve peaks due to the presence of spin-orbit interactions. We hope that our predictions will encourage the experimental realization of suspended carbon nanotubes contacted to ferromagnetic materials to detect the spin, orbital and Kramers polarization and their effect in the SU(4) Kondo effect. 

\begin{acknowledgments} 
We thank helpful discussions with R. Aguado and Rok Zitko. We
acknowledge financial support from Spanish MICINN Grant No. FIS2008-00781 and
the Conselleria d'Innovaci\'o (Govern  de les Illes Baleares, Spain).
\end{acknowledgments}


\begin{thebibliography} {99}
\bibitem{noteSO}
Notice that we have neglected the spin-orbit  contribution responsible for the different spin-orbit gaps observed for electrons and holes that depends on the chiral angle $\Delta_0\propto \cos\theta$. This term is diagonal in the sublattice space diagonal $\Delta_0 s_3\tau_3$ and it gives just a shift in the energy spectrum.
\bibitem{note}
Since $B_{\rm exc}$ depends on the applied external field,  it is possible to find, for a given level position, multiple values for the external field that cancel the same $B_{\rm exc}$.
\end{thebibliography}
\end{document}